# Replications in quantitative and qualitative methods:

# a new era for commensurable digital social sciences

**Dominique Boullier, EPFL, Digital Humanities Institute,**

**Working paper to be published by the University of Siegen (D)[1]**

**November 2018.**

**Abstract**

*Social sciences were built from comparison methods assembling field works and data, either quantitative or qualitative. Big Data offers new opportunities to extend this requirement to build commensurable data sets. The paper tells the story of the two previous quantification eras (census and polls) in order to demonstrate the need for a new agency to be considered as the target of this new generation of social sciences: that of objects as ANT proposed and of replications that propagate all over the digital networks. The case study of the topofil of Boa Vista is revisited to explore how a qualitative method dedicated to comparison and an ANT approach extended in a replication theory may offer new insights from any field study and may use the digital ressources to do so.*

It is widely acknowledged that our times of digital financial capitalism generate a general requirement for evaluation, mobilisation and traceability and that the digital platforms are a key feature of this new experience framework. Consequences extend from quantified-self to high frequency trading, from reputation scores to the new spring of artificial intelligence. We would like to emphasize two dimensions of this world of digital traces made available by the web and the platforms:

- social sciences get affected as much as any other field and will have to revise a large part of their methods, provided that the analysis of the current changes does not miss the important point of traceability ;

- if traces become the main resource, qualitative methods must be transformed as well as quantitative ones, since it is a whole new standpoint in research that is made operational, the one of replications, of propagation of cultural elements.

We shall sum up the historical landmarks in quantification history that made emerge specific standpoints in social sciences. The first tow eras of quantification, censuses and polls, manage to create a sense of reality for society and then for opinion, through carefuly designed statistical processes. The purposes were different, the stakeholders as well but it produced some reflexivity to social groups, to think themselves as societies or as opinions with comparisons made available throughout a nation state, a public or even at the international level. A third era is already taking its role to provide a new style of reflexivity. Big Data, Machine Learning and digital platforms make available a high frequency and finely grained feedback based on digital traces from very low level (e.g. clicks and cookies) to linguistic expressions (e.g. comments and posts). Social sciences have to move towards a smart use of these new sources, without quitting the previous ones, but designing the right understanding of what entities should be the focus of the investigation in this data deluge. We shall argue for a general theory of replications based on digital traces, which means reconsidering some old traditions that were forgotten, such as Tarde's imitation theory, and reformulating the real contribution of Actor Network Theory observations to this new framework: objects, signs and circulating entities have an agency of their own and cannot be crushed under the powerful agencies of structures (societies as a whole) nor of individual preferences (opinions and decision of rational agents). We shall revisit a famous paper by Bruno Latour (1999), « the pedofil/topofil of Boa Vista », to show how new requirements for qualitative field studies can now become a standard, expecially within the actor-network analysis framework. This qualitative methods, focusing on traces, can then become more coherent with some quantitative ones, using massive digital traces, provided that their relevance is limited to some propagation processes, and that the general framework of replications make them comparable and even computable.

---

[1] Many thanks to Jörg Potthast and to Susann Wagenknecht for their comments and careful reviewing.

# Part 1. Making the social commensurable through quantification: historical landmarks

When social sciences as much as society are enduring such a data deluge, labelling it as a "revolution" is less useful than trying to compare our times to other important historical lanmarks in the history of social sciences. Since we are confronted to the general extension of computation to all areas of social life, we need to understand first where these traditions of quantification come from, even though we will limit our investigation to the main changes in large social quantification processes that shape the way social sciences are. Of course, qualitative methods played a role in founding the basis of social sciences:

however, their connexion to large institutions that can reuse them for their own purposes was not so strong as it was for quantitative methods. To say the thruth, it worked the other way: states developing large census reused by social sciences, and media generating opinion polls that were exploited by academics for their own purposes. Sounds familiar, isn't it? Social sciences and data scientists are craving for the huge amount of traces available on digital platforms that are not really eager to reveal what is their main source of revenue! Deriving from this quantitative chronology, we will turn, in the next part, to the challenges of traces availability for qualitative approaches.

## 1.1. How « society » got its agency

If we want to understand the historical times that we are living in as regards quantification methods (and the social reflexivity that is attached to it), we need to look back to the times of the construction of this entity, 'society'. Let us pretend here that Durkheim succeeded in making 'Society' exist. The term was not coined by Durkheim, obviously, although its history is not a long one. The archaeology of the concept of society (Latour, 2005) should be further enriched by calling upon the work of Quetelet, who produced the 'average man' which long remained the key to all statistics. At the end of the nineteenth century, however, and largely thanks to Durkheim's genius, 'society' took a strong stance regarding 'community', which was still prevalent (see Tönnies, 1887). Durkheim's early work on the 'division of labour in society' (1893) was not based on statistical methods, but instead laid the foundation for a model of social types, aggregated in mechanical and organic solidarity. Detailed examination of legal systems served as demonstrations and therefore relied on the groups formed or being formed that are legal systems in their traditional or more modern aspects. With 'The Suicide' (1897), the method was set up to extend the discussion of the types that would reveal anomie to be a problematic situation. But reliance on data records produced by states, from their various components (ministries, prefectures, governments) became key to the demonstration of society's influence on individual behaviours. It was these aggregates that are explained or explanatory, using a method of comparison between countries, regions, counties or districts, where possible and necessary. The method depended entirely on the data available and could not afford to criticise or to question the procedures for the production of this data, despite the countless limitations identified upon publication (Douglas, 1967). By organising his entire systems of proof around these national administrative statistics, Durkheim found a quantitative analogue for his conceptual choice that put 'Society' in a separate status from all manifestations and individual behaviours. Durkheim's *whole* became an entity of the second level, 'Society', (Latour, 2005), while the censuses and other state-data-registers simply perform the task of recovering individual, administrative events (marital status, judicial procedures, etc.), formatted in identical categories and aggregated to reveal the behaviour of populations. Durkheim's strength of conviction was to make these statistics exist as equivalent to his *society:* quantification is double-sided, it is able to account for a *whole* through the quality of exhaustiveness, and at the same time, itaccounts for the agency of the social structure as such, even though the two functions are not necessarily tied together (exhaustiveness may not generate an agency).

It is necessary to note that a *convention* (Eymard-Duvernay and al., 2004) was formed between data producers from the state administrations and the emerging social sciences. Together they produced the entity *society*as the object to be tracked by the state for the purpose of governing and to be explained for scientific reasons. The result is the widely shared and obvious fact that *society* exists, and the methods that allow it to do so have no grounds to be questioned because they demonstrate both their scientific and their operational value: they are 'tools of proof' and 'tools of government' as Desrosières (2014) put it.

Other historical correlations are noteworthy, that do not mean causality but that do allow for an understanding of the power-gains this approach affords in making society exist. In 1890, Hollerith used a machine, a mecanographic calculator, that he had invented a few years earlier (and for which he filed a patent application in 1886) to conduct the U.S. census. The Census Bureau had not yet finished processing the previous census dating back to 1880 when it had had to start the next one. A change in technique was both necessary and available. Hollerith's tabulating calculation-machine did the work and was sold for doing censuses in several countries. His company would later be transformed into IBM by Watson, in 1926. We can see how the power gained in the counting and description of populations reinforces the status of the State and offers it supposedly useful sources of information for its governance. The pretence of the calculation's exhaustiveness seemed to fulfil the promise of the concept of society: a technical device capable of inputting all that existed, that as Hollerith's census-procedure-equipping machine

In modern Western countries, the State gained its legitimacy through electoral processes that rely on "nations", those "imagined communities" (Anderson, 1991) that work as a contenant ( a value, a concept) transmitted to the minds of the population while the State is the container (Boullier, 2010), in control of the bodies of the people, of their individual physical behaviour (borders, location) or as the 'materiality' and the 'statement' that constitute the apparatus in Foucaldian terms (Foucault, 1982). This is why 'society' is always enacted in various 'nations' although social theory tries to extract it from the limitations of national boundaries. The first generation of social sciences was indeed doomed to methodological nationalism (Beck, Sassen) and still has problems inventing the methods to account for a globalised world made of flows (finance, media, commodities, migrants, and so on). However, we shall see that when examining this digital world, it is quite difficult for us to escape 'methodological platformism', due to the total dependency on the traces data platforms deliver!

## *1.2 Opinion exists, I invented it[2]*

After the first generation of quantification, used by Durkheim to build the concept of *society*, we could give the label '2G' to the emergence of *public opinion* in the late 1930s. In 1936 Gallup was able to predict the election of Roosevelt over Landon, based on a survey of 50,000 people. Roper and Crossley had done likewise at the same time. Gallup not only impressed the media and policy makers, he radically disqualified older methods (straw polls), including that of the Literary Digest, based on responses from 2 million people, whilst even predicting their own erroneous results (Osborne and Rose, 1999). This impressive demonstration lays the foundations of the survey's reliability and of investigative sampling methods. The exhaustiveness of inquiries on entire populations was indeed sacrificed in the process, but the new approach managed to produce accurate results, provided that the terms of *representativeness* were respected. It nevertheless failed to predict the victory of Truman in 1948, whose voters changed their minds in the last ten days. Methods thus applied to political life and to life-size tests as important as a presidential election had previously been tested on readership studies for which Gallup had operationalised stratified sampling. In fact, these methods had already been applied by the Norwegian Kiaer in 1894.

---

[2] The works of Osborne and Rose (1999) and Blondiaux (1998) develop this story extensively.

Similar statistical methods in the field of agriculture and later in unemployment in the early 1930s, in the USA, underwent profound changes, from the correspondents' method to random sampling based on probabilistic approaches (Desrosières, 1998, Didier, 2002). Quota methods based on 'sensible choices', where the selected sample is matched with certain properties of the population identified by the census, were however different from those methods of stratified random selection, and were even despised somewhat by statisticians (cf. Stephan quoted by Didier, 2009). The data collected were also very different, since statisticians from agricultural or employment administrations wanted to obtain 'facts', but were nevertheless obliged to rely on statements, not measuring machines, even if they attempted the latter with the 'crop meters'. Yet the sampling-legitimisation-operation generally succeeded, primarily thanks to Gallup's (1939) performances, which were dedicated entirely to other social worlds; those of 'public opinion' and not 'society'. The latter remained a reference of statisticians of the federal state and its offices. It was unquestionably in the context of the mass media that the importance of sampling was recognised. With Ogilvy, Gallup studied film audiences, and then with Crossley, at Young's and Rubicam's consultant firms, he studied radio audiences, using telephone interviews before even making a proposal to conduct the election polls. From this point of view, Gallup found his counterpart in academiawith Lazarsfeld, who in the same period, in 1936, launched a "Radio Research Program", based on audience-research-work begun in 1930. Together with Merton they launched the *focus groups* method as early as 1941, and their study of the city of Decatur in 1945 provided the data for the analysis of "Personal Influence" published in 1955 (Katz and Lazarsfeld, 1955). The latter study established the framework for analysis of the *two-step flow* : mass media play a role but do nor act as a general and direct flow to people's minds, despite their "broadcasting" technology but through the mediation of individuals who act as*influentials* .

The links between the mass media and politics are thus elements of new statistical methods. As Alain Desrosières noted (1998), a national election's predictability actually depended upon the formation of a common public media-space across the United States, and only the radio could do this in such a way as to make the voter's knowledge about electoral candidates comparable. Considerable media transformation and the mass media (radio at the time) established the conditions for the emergence and validation of a survey technique, which thus opens up a whole new era, most notably for political science and market research. Moreover, it is *public opinion* itself which takes on a measurable existence with these sampling methods whose performative power will by far exceed their experimental phase. This move does not disqualify the previous alignment made by censuses and mecanographic calculators that will soon be transformed in computers. But the role of radio (and telephone for the report of results and their aggregation at a faster speed) contributes to a new apparatus.

The missing link in my description remains the vehicle of financial incentives for such investments, needed to understand a public. Communication agencies such as polling organisations cannot live solely from their campaign activities, even if they do bring them high visibility and renown. From the outset, their target was the mass media, as noted above, for one essential reason: audience measurement has been the key to the distribution of advertising space, since the dawn of radio and then later with television (in 1941 the first advertisements were aired on American television for Bulova watches, during a baseball game). But these measures also serve to monitor the impacts of these campaigns on the minds of consumers, giving an unprecedented boost to marketing, which in turn drives increasingly sophisticated communication strategies (Cochoy, 1999). Brands have thus been present, from the outset, in methods of inquiry into opinion using sampling; that is, from the moment such investigations were aimed primarily at mass-media audiences. Market research on consumer goods developed at the same time, from the 1930s, and in the same movement of national standardisation of products, as Desrosières pointed out. The production of a unified national territory, through the media, that included transportation and mail, established a new condition of felicity for these survey methods. This allows me to draw a direct parallel with the recent creation of a global market, this time beyond the national ones, through the domination of digital platforms. Google, Apple, Facebook, Amazon and Microsoft have produced the same effect on a global territorial scale as radio and the railway had on the territory of national markets. This is in line with the work of McLuhan (1964), for whom the change of scale in itself constitutes another kind of collective experiencefar beyond the features of the contents exchanged.

The work done by Gallup on the business side and Lazarsfeld on the scientific side provides whole societies with methods with which to analyse themselves and to represent themselves – as opinions. Tarde (1901) has certainly paved the way for understanding the formation of publics and opinion, yet it is only when the metrics are established and produced in a conventional way that opinion finally exists. Only the media's control and their ability to produce a unified public in a national territory enabled this methodological assemblage to hold. The 'whole' referred to by the polls is in fact originally the *public* formed by the media, which allow the *audience* to emerge as *public opinion* and to make it permanently visible and measurable. At this point, everyone knows that 'opinion exists', whatever the report about the artefacts needed to make it exist, and despite what Bourdieu (1984) said about it ("Public opinion does not exist")). Opinion has been naturalised, 'taken for granted', and the sampling methods lie buried beneath the powerful performative effect of these immediate, aggregated indices, thse scores used by media for ranking preferences about politicians or products. The approximation that is measured through confidence intervals is made acceptable, especially with the repetition of the same questionnaire over time (by panel, independent rotating sample) under identical conditions, 'all things being equal'. It allows for the smoothing-out of biases which then become acceptable by convention.

## *1.3.* *The era of digital traces make a theory of replications possible*

The two first stages of quantification in nation-states history cannot avoid the digital transformation and they have been usingomputing power to amplify their own way of calculation. However, the collecting devices are now designed in networks and the traditional databases (registar, polling populations, etc.) are exceeded by the data deluge rising from the traces of behaviour available at a low cost on any digital platform. No official record, nor personally controlled expression by pollsters but traces of behaviour, some of them linguistic (comments, posts, tweets, etc.)[3], but most of them very low-level traces, such as clicks, time of exposure, likes, and so on. All of them are collected, produced, analyzed, recorded, sold by digital platforms ranging from the GAFAM (Google, Apple, Facebook, Amazon, Microsoft) to smaller ones in cultural contents (Spotify, Netflix), transportation (Uber) and services of all kinds. Personal data and privacy are not the only issue in these environments because profiles and patterns are elaborated through clusters of features that may not be related at all to the traditional socio-demographic features that made the previous quantification eras so efficient for government and scientific purposes at the same time.

While the statistical requirements for data quality (exhaustiveness and representativeness) can no longer be applied to dynamic universes such as social networks and the web, Big Data has produced ersatz that are volume and variety. However, another feature emerged, velocity (which thus constitutes the 3V of Big Data) which had no equivalent in the social sciences and in the quantification of the social. We have proposed (2017) to make traceability a statistical requirement as such yet to be built. It makes it possible to account for new entities such as *replication,* these contents which circulate, and which until now had hardly any place in *society* (calculated thanks to the use of the exhaustiveness of registers) and *opinion* (calculated thanks to the construction of representativeness via surveys). This universe is more familiar to the reader because it is the digital world of our daily experience. We will limit ourselves here to presenting a systematic table for the long-term comparison between periods.

---

[3] The fact that these expressions are linguistic make us believe they look like *opinion* (Boullier and Lohard, 2012) which a real misunderstanding about how public opinion gain visibility, neglecting the key role of polling institutions in frmaing these expressions and making them comparable and computable. This is precisely used as an argument for those who praise the "naturalism" and the spontaneity of these expressions on the web or on social networks to challenge the artefact of opinion polls. These statements ignore the framing process of platforms and the need for a conventional metrics to make opinion exists. What we obtain when scraping the expressions on the web, is a new kind of traces, that have to thought about as such, and this is why we propose this theory of replications that will delineate the third generation of social sciences.

| | 1st generation | 2nd generation | 3rd generation |
|---|---|---|---|
| **Concept of the social** | **Society/(ies)** | **Opinion(s)** | **Replication(s)** |
| *Collection devices* | Censuses | Surveys/Polls | Platforms |
| *Statistical requirements* | Exhaustiveness | Representativeness | Traceability |
| *Co-construction institutions/ research* | Registers/ inquiries | Audience/ Polls | Traces/ Repurposed digital methods |
| *Major players of reference (and funding)* | States | Mass media | Platforms |
| *Operational actors* | National Institutes | Polling organisations | Web platforms (GAFAM) |
| *Founding authors* | Durkheim | Gallup, Lazarsfeld | Callon, Latour, Law |
| *Key problems of scientific approaches* | Division of labour and the welfare state | Propaganda and media-influence (audience measurement) | Science and technology (scientometrics) |
| *Technical conditions* | Hollerith's machine (tabulating calculation) | Radio and telephone | Internet, Web and Big Data |
| *Technical criteria for data quality* | Relevance, accuracy, timeliness, accessibility, comparability, coherence | Confidence intervals  Probabilities | Volume, Variety and Velocity |
| *The social science's dominant modalities* | Explanations | Descriptive and predictive correlations | Predictive correlations, memetics |

## *1.4.* *From genealogy to standpoints*

This historical inventory that we have formulated in terms of generations because technical and methodical eras correspond to it, appears to us to reflect indeed a diversity of approaches to the social that have coexisted at the very start of social sciences. The division of disciplines (notably between sociology, economics and psychology) as well as the absence of reliable empirical methods for certain approaches only make it possible to explain the hegemonic claim of each approach that led to this sometimes aggressive division of labor. Tarde, an ancestor who carried the approach of traces and propagations, never enjoyed academic recognition comparable to that of Durkheim or Weber. It is not the digital technology that makes these new entities emerge, it only makes them visible and calculable. They were already part of ordinary experience (like conversations and rumours, Boullier, 2004) and had been conceptualized by Tarde (1890) in his theory of imitation (which included opposition and invention) but were not traceable nor computable. It is therefore preferable to propose a synchronic table of these approaches.

| Generations | 1st Society | 2nd Opinion | 3rd Replications |
|---|---|---|---|
| **Standpoints** | **Structure** | **Individual preferences** | **Propagation** |
| **Wavelengths** | Long term | Mid-term , cycles | High frequency |
| **Features of the networks** | Structure | Nodes | Circulating entities |
| **Main concern** | Positions | Decision | Propagation |
| **Process** | Inheritance | Arbitrage | Neighbourhood |
| **Status of human actors** | Inheritors, determined subjects | Strategist, decision-maker, rational agent | Vehicle for memes'propagation |

Each of them works as a standpoint on the world. Each point of view depends entirely on the instruments it mobilizes and which frame the world and allow its capture at the same time (Latour, 1987). But these standpoints focus on particular entities that in each case have a privileged status that we will have to characterize. They encompass methodological choices that cannot be used for other purposes. Admitting this limitation prevents from any temptation to imagine any God's eye view that would assemble all viewpoints at the same time. The focus on "structures", associated with the exhaustiveness of censuses, clearly differs from the focus on "individual preferences" associated with the representativeness of surveys that are used primarily in marketing, in public opinion surveys but also in all major attitude surveys used by economists, for example on the feeling of happiness or confidence in the future, etc. These two entities (social structure and individual preferences[4] aggregated under the so-called « market ») are themselves very different from circulating entities which can be messages, goods or beings which have their specific properties analysed sometimes in semiotics but taken seriously by memetics (Dawkins, 1976, Blackmore, 1999, Dennett, 2017).

The best examples are found in propagation patterns of tweets or memes (Boullier, 2018). Memes are "elements of culture" (Dennett, 2017) that spread across human minds and artifacts of memory, "representational states that propagate through representational media" as Hutchins labelled them. On the web , memes became famous by the high speed replications processes designed on sites such as 9gag.com or knowyourmeme.com. Catchphrases, slogans, stereotyped images, buzzwords and so on are the perfect candidates for testing the agency of these messages that are highly contagious, despite people's will or taste. However, more complex items can circulate as well, including political ideas or fashion trends and the contagion process can include phases of derivation, where the content is slightly transformed and manage to reach a new audience through this transformation. Some analogy with translation processes can be found and sociolinguistic influences (loanwords for instance). The theory of replications that we want to build is based on these processes and is not limited to the restricted sense of memes (on the web or in the atomistic definition of Dawkins) nor to the virality process because derivations occur and must be accounted for. Twitter should be considered as the perfect replication machine, since the invention of the retweet button in 2011, because it creates a nudge for users to replicate the tweet without any comment, at a high frequency rate. We made a comparative analysis of more than 1000 data sciences papers on Twitter in the years 2016 and 2017 that show the prevalence of traditional distribution of agency to understand propagation processes: either the structure of the network (degrees) or the role of specific influentials are considered as the main variables to account for the spread and the recurrence of tweets. However, about 20% of the papers adopt a different view and distribute the agency, by hypothesis, to the features of the tweets. This has been well publicized in the case of "fake news" propagation: the large spread of fake news was neither correlated to the structure of the network nor to the role of influential nor to the intrinsic falseness of the content, but to the novelty score. This feature is an expression of the competition for attention among memes: the only fact of chocking some audience by a fresh and striking news is enough to facilitate its diffusion in a social network. This does not mean that the structures of the network or the preferences of some individuals acting as influential do not play their part. But a specific agency can be distributed to the features of the "fake news". The digital mechanism of replication is becoming so powerful for collective minds that we consider Twitter as the drosophila of these third generation of social sciences.That is to say that even though Twitter is clearly *not* an expression of the "real life" (as much as the drosophia was clearly not a reduced model of human genetics), the propagation of memes within this milieu is a perfect field of experimentation for replication processes, provided that they are processed at the adequate formal level.

---

[4] Individual preferences are the focus of psychology and economics. « Social psychology » tries to extend the realm of individual preferences to social groups and collective effects (such as in network analysis). However, this aggregative trend is much more salient and powerful in economics that built the concept of the « market » to account for the coordination of all individual preferences. It looked so powerful, as is public opinion, because the theory of the « market » looked like an layman experience as one can do in market places. However the concept built a reality of its own, because its laws that classical economics designed (from Adam Smith to Walras), including the « invisible hand » agency, managed to make the market operational, as in a performative move, to the point that it is supposed to dictate all policies. This is well described by K. Polanyi in « The Great Transformation » who tells the story of how the economy got disembedded from society during the 19th century .

Let us take classic examples from the sociological literature to better understand the scope of this "three standpoints" framework even though it is only in our digital times that it becomes possible to trace this processes: one example from Durkheim's suicide, another one about Mauss' gift, another one from Actor Network Theory and the final one from social network analysisSome will give the insights about the need for a replication theory, others, on the contrary, will demonstrate their reluctance to do so and their ability to act as if one specific standpoint can account for every feature in a given situation.

a/ The classic opposition between Durkheim and Tarde makes sense in this context. When Durkheim studies suicide, he does not enter into an analysis of choice by individuals even though his categories may appear psychologically significant. Selfish, altruistic or anomic suicides all depend on the pressure that society, as a social structure, exerts on these individuals, to varying degrees (the degree of social integration). (revoir ici ce qu'en dit Boudon)The statistical methods used by Durkheim have been considered by the scientific community as sufficient to show these structural effects in the long term. They do not deal with processes of influence, beliefs or values, for example, that Weber would have used to explain choices and preferences for suicide (voir Boudon). These methods also do not capture processes of rapid, high frequency and much less "chosen" contagion as can be observed in some situations of suicide spread, in public, in companies or in entire societies as in Micronesia in the 1970s (Rubinstein, 1983). For Durkheim, "the example is the occasional cause that makes the impulse burst ; but it is not he who creates it and, if it did not exist, he would be harmless" (Durkheim, 1897:135). Tarde replies that "all social phenomena, all social influences, such as physical and physiological influences, consist of repetition of similar acts, repetition-wave or repetition-heredity" (Tarde, 1897:41). We understand that the confrontation is irreducible, when in fact, they each study different moments of the social (high frequency and "in the making", or long duration and "ready made") with different points of view on very specific entities that have their own agency.

b/ Even more radical and unsuspected is the opposition between Durkheim and Mauss on the question of the entities that primitive societies make exist, and it will anticipate on the evolution we advocate for qualitative analysis. In his essay on the gift, Mauss (1950) includes Malinowski's work on the kula (1922), ritual exchanges between tribes in the Trobriand Islands. Malinowski saw it as a form of moral obligation to maintain peace between the different groups and to generate economic exchanges (market and strategies). Sahlins (1976), on the other hand, constructed a historical model of types of reciprocity, to highlight a structural effect. But Mauss, for his part, focused on the « hau », on the spirit of the given thing, that is to say on the attributes of the circulating entity, to which he confers a power of obligation because of the "spirit of the given thing" always present in the exchanges. Lévi-Strauss (1950) criticized this abusively indigenous vision according to him, even though he himself based all his theory on exchanges (of goods, women and signs) but on the condition that the structure of these exchanges was freed of any power from the attributes or entities exchanged.

c/ Following Mauss' track  about the power of objects but from a very different origin, Actor-Network Theory founders (Callon, Latour, Law) provoked a tremendous shock when they introduced the role of so-called "non-humans" in the sociological analysis. For instance, Callon compared the way scallops reject domestication by scientists to the one of fishermen against regulations (Callon, 1986).  The opposition seemed to be an ontological issue against the crushing of human/nature or human/artefact distinctions. But the issue was in fact methodological. While ANT restored a form of agency to circulating entities since they could contribute to the constitution of a network or even emerge as an actor-network, structural approaches focused on the effects of position, hierarchy, reproduction, and domination that effectively record the existing asymmetry between humans and objects. This did not invalidate the structural and inheritance approach of positions and attributes, despite the tendency of ANT authors to disqualify it by showing the artificially constructed character of this "dispatcher" (Latour, 2005).

d/ In this broad picture, the methods of social network analysis are supposed to be the most likely to exploit the new digital resources. However they have in fact perpetuated the privileged power to act of (network) structures and individual preferences (nodes, influencers) (cf. Watts and Dodds, 2007). Network topology methods and their mathematization using graph theories reinforce the focus on structure since it is possible to identify clusters, densities and connections, structural holes (Burt 1992), bridges or gatekeepers, all properties that are related to positions in the overall network structure. Other approaches will focus on the role of network nodes, including dealing with "personal networks" and stars of varying degrees (Barnes, 1972). Nodes then refer to individuals whose properties can be measured through their centrality scores but whose influence (Katz and Lazarsfeld, 1955) on the behaviour of other network members can also be identified. This distribution of power to act, of agency, was quite well identified by Granovetter in his famous paper on "the Strength of Weak Ties" : "Some of the studies have emphsized the ways in which behaviour is shaped and constrained by one's network (Bott, Mayer, Frankenberg), others the ways in which individuals can manipulate these networks to achieve specific goals (Mayer, Boissevain, Kapferer)". No reference at all to any agency of the news, messages, ideas that circulate in these two models even though Katz Lazarsfeld were careful at making distinctions in influencers' role depending on the topic of the conversation.

Pentland's social physics (2014) also mobilizes sociological analysis of networks. It manages to simplify social situations by adopting only two of the traditional social science perspectives. Two factors are identified as predictive of organizational dynamics: a/ "engagement" as a function of social pressure or "social learning" (extracted from indicators of internal connection to social groups) and b/ "exploration" as a function of individual initiative (such as that of a gatekeeper or bridge in social network analyses such as Granovetter's (1973). When it claims to study the flow of ideas, one would expect some consideration for the entities that circulate, an analysis of the contents of the exchanges, the follow-up of specific innovations and their transformations and so on. In fact, he tells us, "the numerical estimate of the flow of ideas corresponds to the proportion of users who are likely to adopt a new idea introduced into the social network. This flow of ideas takes into account all the elements of an influence model: the structure of the network, the strength of social influence and the susceptibility of individuals to new ideas" (2014: 83-84). In these last three agencies, no place is left for the role payed by ideas themselves , their design, their presentation, their capacity to capture attention through semiotics saliences, and so on. The networks analysis seem to proceed exactly the same whether people exchange jokes, scientific information, spare parts or cabbages! That is why we must avoid considering networks as the key word of any digital sociology, and rather develop a theory of replication and circulating entities where their features will play their role.

What lessons can we learn from this foundging works (to which scientometry should have been added to be complete)? From Mauss (and his mana) to ANT (and his nonhumans), conceptual frameworks emerge to think of the agency (or « power to act ») of entities other than social structures or individual preferences, while for Durkheim and social networks analysts even though the observations should trigger an investigation of the replications, they maintain their initial standpoint.

The entities observed (structure, preferences or replications) constitute entry points into the social world which will, inevitably, remain their exit points. eThe agency distribution is also done in qualitative approaches. The framing of any topic, of any social issue generates a whole set of methods, of entities, of agencies, that are completely different depending on the adopted standpoint, even though, while using one specific method, the scholar is not really aware of the framing that they encapsulate. Let us compare some issues to see how the only fact of selecting a question to address will require specific methods because of the standpoint that is encapsulated within both the issue and the method. Studying a crisis (for example, a crowd of supporters in a stadium, as we did in Boullier, Chevrier, Juguet, 2012) will not mobilize the same methods as, for example, the social distribution of higher education students according to their fields of study. The entities observed are not the same, the rhythm of the processes is unrelated (the high frequency and short intensity versus the long duration of reproduction of inherited social positions). But beyond that, the questions asked and the issues are not of the same nature either.  The use of interviews and questionaires wouod introduce us to the representations and preferences of individuals that are more adapted to analyse the culture, the discourses, and not the structural effects. We can certainly study the long-term social determinants of hooliganism or the social composition of the stands at the Marseilles velodrome (Bromberger, 1995), and statistical analysis of cohorts of supporters would be fine for that (structure agency). But this will be of little use in understanding how a riot is triggered at that precise moment in time because one needs to collect very finely grained observations of behaviours on site (and not social positions nor opinions) and conversations on line, connected to the events on the field of the game for instance.

Braudel had stressed the importance of these different points of view on social time: "a clear awareness of this plurality of social time is indispensable to a common methodology of the human sciences" (1958: 726). He had thus distinguished events, cycles and long duration by making it clear that "the only mistake, in my opinion, would be to choose one of these histories to the exclusion of the others" (1958:734). What Braudel expressed is a kind of precaution principle that commands not to rely on one and only one standpoint to explain all kinds of social times. However, the methods that are used for each of these histories cannot be the same whether one focuses on "events, cycles or long duration".

Another way of expressing these different standpoints while not discarding any of them would be to speak of inheritance/ arbitrage and neighbourhood (« héritage/ arbitrage/ voisinage » in french) (Boullier, 2010)

We would say that the structures are part of a process of *inheritance*, well highlighted by Bourdieu (1979). There is no point in contesting or flattening them: the social heritage has its own « laws » that have a structural effect, that acts as a "ballast" for our behaviour, even if it is userful trace the mediations by which it is propagated. Individual preferences, choices and decisions are moments of *arbitrage* that weigh possible influences, which are in constant conflict. But these moments of decision, which are often attributed to a rational, calculating or strategic agent, occur only in certain situations, without invalidating the effects of inheritances and neighbourhoods.. Agents' intentions, strategies and inner states of mind can then be part of the conceptual framework in the same way as economics or cognitive psychology do to account for these preferences. Replications are a matter of *neighbourhood*, of imitations by proximity, by crowd or public movements but always thanks to the specific power of what circulates: a passion, a desire, a belief, an idea, a message, a meme, the « hau », etc. It is because these entities have this power to act, this agency, that they provoke these emergences that cannot be reduced to the inheritances of structures, even though many parameters of the situation may be affected. But the problems raised are then totally different because of this conceptual, methodological framework, which constitutes *one* standpoint among the three available. Inheritance, arbitrage and neighbourhood are therefore in no way contradictory but cannot be captured together because the methods for reporting on them are radically different. Digital technology does not change this requirement for a standpoint, but amplifies one of them (replications) more than the others, to the point of finally making accessible a specific section of society, that of replication, made of virality and memetics, of propagations and vibrations, thanks to the amplification of the traces, that make visible the agency of these circulating entities.

tThese points of view are not limited to quantitative approaches, whereas all questions of scientific method face the same problems, whatever the survey techniques they use, and must also adopt one of the standpoints mentioned. We will see how the third point of view, that of replications and circulating entities, can be extracted from one STS field observation, that is designed for accounting for the propagation of "scientific reference" through a number of devices, of data formats, in order to move back and forth from "the world" to scientific knowledge". Actor-network theory (ANT) is the framework used in the paper we discuss but the questions we want to raise would the same for any other qualitative field work. However, the theoretical framework of replication would lead to important revisions of the ANT methods that we want to highlight here.

# Part 2: Making the social commensurable through qualitative comparison: re-inventing ANT

Bruno Latour's "Topofil of Boa Vista" will serve us to illustrate how one of the best qualitative field observation can be reframed in terms of replication and what it requires to make comparison possible in order to deliver more controlled statements. One of the main requirements will be to more clearly select the standpoint from which the observation is made, because the delimitation of the field, the entities that come under scrutiny, the methods to collect and compare data will be different. The story is told by an uthor who manages, as often, to create a plot from plot a research agenda involving pedologists (soil specialists) and botanists (specialists in equatorial vegetation). They have one question in common, a challenge not only scientific but political and historical since what is at stake in the undestanding of the forest changes: is it the forest that advances on the savannah or the opposite in this part of the Amazonian forest, specially chosen to test the two hypotheses ? This way of framing the purpose of the observation is a good way to captivate attention but it is also in line with what ANT emphasize in sciences and technology studies: "science in the making" is always embedded in controversies, is always debattable before a "fact" can be considered as "ready-made". Fact have to be well fabricated before it becomes quasi natural. This paper is one of the many examples of the field works done by Bruno Latour within scientific labs, the most famous one being the first one that was told in a book called "laboratory life" (with Steve Woolgar in 1979) that paves the way for a complete shift in the way scientific activity is understood by social sciences. The connexion with technology and innovation in ANT's works is made through the new role devoted to devices, traces, material settings, what was called "non-humans". In this paper, the same focus on devices (such as the pedofil) will be the common thread that will populate the story with realistic notes. Photographs are widely used to render this immersion feeling for the reader and to produce evidence for the scientific demonstration. There are countless articles in STS and ANT that make an object, a device, a technical system the hero of the sociological narrative: the Berlin key (Latour, 1993), the electric vehicle (Callon, 1979), Aramis (Latour, 1996), the bicycle (Pinch and Bijker, 1984), the air pump (Shapin and Shaffer, 1985) , the stocks ticker (Preda, 2006), the Black-Scholes model (Mc Kenzie, 2006), etc. And this immediately appears to the reader: the narrative is populated with living objects or entities, such as the forest, that would never have had their place in observations oriented towards the agency of structure or individual preferences. Bruno Latour ( 1996) advocated an "interobjectivity" that will dissolve the usual and simplistic distribution of status to objects: slaves, masters or substrates of a sign. When Durkheim discussed the role of the flag in his work on religion, he tries carefully not to deliver any agency to the flag, except that of a slave of human passions or imagination. In thelist of objects like linoneum or perfumes that Bourdieu mentions in his "Distinction", they convey the dispositions that the individual will incorporate but they are just slaves of the social positions, adding nothing of their own.  They remain in their place, that of *slaves*,. When they are lucky enough to be taken seriously, in the form of technical systems, such as the steam engine for Marx or the railways for Proudhon, they very quickly tend to become *masters*, in the form of a technical determinism, a fatal agency of the production forces, even if they are enshrined in production relationships for Marx or found as alternatives for Proudhon who politicizes these architectural choices in a vein that echoes current demands for technical democracy. When Barthes (1957) describes objects in short notes that have become famous as his « Citroën DS » or in more analytical essays such as the fashion system (1967), it is their value as a *sign* that is highlighted, which quickly slips towards the status of « lieu-tenant » (substitute)  of something else, of a cultural, political or even civilizational stake.

## *2.1. The chain of reference: devices AND statements*

The topofil of Boa Vista is neither omnipotent nor impotent; it will prove to be a decisive mediator, a link in the chain of propagation of the "quasi facts" not yet made during this experiment. It is a mediation because it translates the state of the world in another register, in another material format but also a significant one: a forest becomes, thanks to its systematically unwound thread, a laboratory that grids the ground and makes possible an organized, controlled collection of very heterogeneous data, coordinated between different disciplines. The topofil, however, acted only for its part, because no one would deny that researchers made decisions or that the framework inherited from scientific experiments on the field and their protocols was also an essential reference. Certainly, but without the topofil, none of these agents, supposed to be capable of acting alone, could have been coordinated as a precarious laboratory in such a forest. It has not only transported a grid model, it has produced a new material framework and allowed intelligibility between researchers, it has made this small piece of the world *commensurable*. The somewhat teasing or very rigorous reader might wonder why such importance is given to this object when, as the article goes on to show, a quantity of other objects are needed to produce a scientific article. Indeed, the cupboardwhere the botanist keeps her plant samples is just as important. The display of the whole chain of mediations is a very useful summary of the whole analysis and will help the reader memorize all the steps and transformations required for a « fact » to emerge, a chain that can be represented as below (revised from a schema present in the paper).

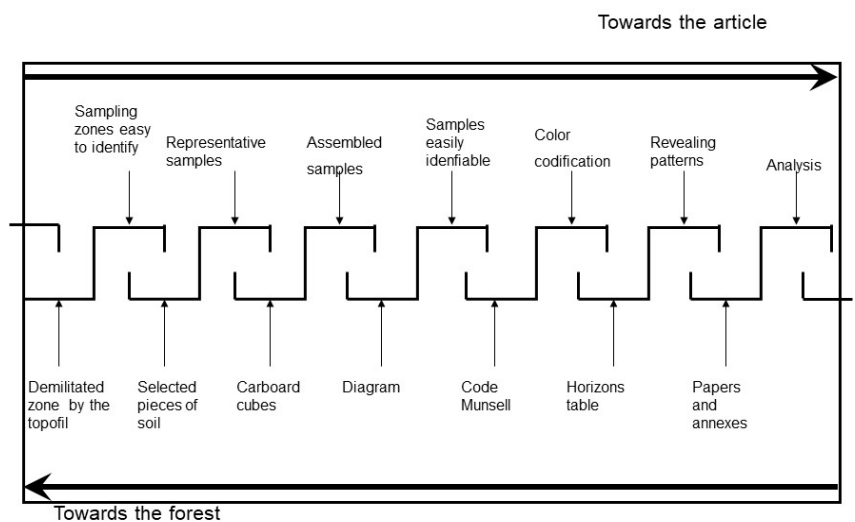

The main conceptual innovation in ANT's studies of science is this very well documented process that produces knowledge : any epistemologic debate comes to be empirically described through the methods, the equipment, the devices and so on. However, one should notice that this chain, even though it looks quite long and complex, is still a *reduction* proposed by the author and since there is nothing like a repository of the original dataset, the reader is compelled to accept or reject it without any further refinement. We are not aware of any validation process from the scientists that made the experiment, which might be an issue when ANT requires to follow the actors… but not to let them intrude in the social scientist's account ! The added value of such an extended description is not to pretend to be complete, to account for all the entities, processes, material that contributed to the research : this pretention to some kind of a « whole », even though deployed in detail, will lead to obliterating the necessary work of reduction for making scientific statements comparable. . In order to obtain a more accountable and comparable report, the ANT style of description should 1/ accept the reductionist effect that any observation is doomed to produce and 2/ shift the focus within the chain of activities from the agency of objects, to the one of the traces that these devices propagate into the chain of reference. At some moments, the paper acccounts for these transformations, for instance when a dataset about the soil is translated into a visual diagram that will be integrated into the final scientific paper. For this purpose, the observation techniques may benefit from the digital traceability of all fragments of activity. As we will show in our revised protocol, by collecting the traces of all messages and data that circulate from one device to another, within the setting of the fieldwork, we could put the emphasis on the agency of these cognitive ressources, on their propagation skills until the final scientific paper. Moreover, since these traces (from one digitized device to another) are easily assembled in data bases and dsiplayed in timelines, we could improve the comparability of these flows of signals (between fieldworks or scientific teams for instance) and even compute these traces. The storytelling of ANT observations was powerful in delivering insights on objects and processes that were discarded by structural analysis or preferences frameworks. However the digital traceability of many entities can improve the precision of the accounts and make them comparable, and so doing  more debattable, and maybe even computable.

 In order to show how one can move from story to traceability, Let us take the example of another device, that could be considered as decisive as the topofil. The pedocomparator that allows "at a glance" to categorize a soil sample taken from a square bounded by the topofil, from its color, using the Munsell code, a largely shared and conventional register of colours. Not by doing a big calculation or by going back to manuals, but only by passing the piece of soil under the holes of the pedocomparator which are surrounded by a color that one can "immediately" compare to that of the sample to classify it according to the standards of the discipline. The magic off well-designed objects requires to incorporate a long chain of mediations and totake into account our limited cognitive capacities, to generate « affordances » (Gibson) or « prises » (Bessy and Chateauraynaud) enacted at a level of perception that does not require verbalization. A minimum of cognitive work for a maximum effect of knowledge, that Hutchins (1995) demonstrated about the way in which the American navy's aircraft carriers were plotted, during the fix cycle, thanks to elementary devices requiring no calculation but only controlled movements, translations and propagations. As Hutchins says, « the fix cycle is accomplished by the propoagation of representational states across a series of representational media » (p. 117). Obviously, in Latour's paper, the description of the mediations is focused on the role of « representational media » and less on the propagation of « representational states », even though they appear as the « substance » transmitted by the agency of the objects. This should make us emphasize the two-fold approach in the theory of replications : devices (material, representational media) AND representations (mental, representational states). Both are connected and their agency should be considered as part of the distribution of agency. The local « power to act » of devices must be accounted for by sociographic approaches (and it might become a purpose for comparison among various social settings : does the pedocomparator always works that way ?). However, the messages, the content or the representational states are the ones that circulate and propagate from devices to devices and from minds to minds. The formatting power of these contents on the accounts is just great ( as we see from these scientific careful examination of pieces of messages.

## 2.2. The agency of objects and the agency of circulating entities: from ANT to replications

Admittedly, entry through objects is the most attractive, the most captivating for the reader who can easily understand that the topofil, the binder or the pedocomparator do something, transform and translate. However, what survives, what will modify the conditions of action of the next link, is the information itself, namely the square thus drawn on the ground by the topofil with usable coordinates in a geographical reference frame, the color indexed from the pedocomparator according to the Munsell code which can then move in the form of a number, or the category of the plant thus identifiable by a name or a code number also. Bruno Latour insists on this double dimension of translation devices, hardware and software (a materiality and a statement would have said Foucault (1966) describing the apparatus), even though the statement is on a very small scale, an indicator, a label, etc. It would therefore be a contradiction to attribute all the power to act to the tool (the representational media for Hutchins) used when what will circulate and affect the conditions of action of the following ones in the chain are data, signals, terms or categories (the representational states in Hutchins' terms). A consistent non-human-centered theory based on ANT must emphasize the agency of these circulating entities, as did Tarde, to understand their power to act on individual brains as well as on structures. Therefore, the method of analysis is very important. Tracking each transformation and the tools that make it happen is not enough, despite the pleasure the reader gets from highlighting the role of previously invisible objects.

The chain that Bruno Latour draws in his diagram is therefore the most important element, because through it, in this circulation, the scientific reference is fabricated, which is neither in the world nor in the minds but in this move back and forth, equipped with mediations that each do their work. The propagation of information[5] is certainly not linear, because the links in the chain show some play in their attachment, as the diagram clearly shows: depending on their translation work, online loss is more or less important.

---

[5] In the paper as in any scientific activity, information can appear as tables, curves, diagrams, notes, whatever is needed to translate one stage of exploration ot the next one that will be processed by another device. But in general, information that is processed in replications can range from low-level signals (including perceptual clues in any situation that may act as affordances) tot highly elaborated messsages (texts, images and more).

This is the basis of any theory of replication, which is not simple capture, recording and transmission of traces but slight transformations, interpretations, translations and treasons, these small differences that Leibniz, Tarde or Deleuze valued just as much. The entity that spreads does so only on the condition of transforming itself (« to exist is to differ » as Tarde used to say). And several translations that are too "loose" or on the contrary too "literal" will give different chances of spreading the initial information. Because of these derivations that were quite well emphasized by the founders of memetics (Dawkins, Blackmore) inspired by the life of genes, we speak of replications and not of « contagion » or « virality » in which the viruses are propagated exactly in a similar form. It is then possible to compare the propagation patterns of these similar entities with memes or even to correlate their propagations to certain qualities of the process of translation, chaining, mediation. Some teams working on the same subject as those of Boa Vista may have adopted the same tools, the same protocols and carried out a slightly different translation at one step or another in the process of building knowledge: a badly done control or on the contrary an overly detailed control, a less advanced expertise in drawing schemas, etc., everything can contribute to weaken their chain of conviction, the robustness of their statement.

But this variation, these different results in comparable situations, the reader of the article does not have access to them. Here is where the ANT program could benefit from a more demanding control of the observations. And we shall propose a way to do it that will make the principle of non-hulman agency more robust, more controllable and commensurable. Thepurpose of Latour's paper is to make a point in the philosophy of reference to establish a conceptual framework of general value and not to investigate one of the links in relation to the scientific methods specific to one field. However, testing the author's intuitions, once the fecundity of this new way of thinking about reference has been admitted, would require a systematic comparison protocol, akin to the one we shall propose below. The power to act, the agency, either of the objects or of the memes (data, categories, etc.) that are propagated could then be tested within the limits of well-knowned experimental reproducibility. The observer focused indeed on a specific moment in the development of scientific knowledge, but only testing can help understand why all these links are important. Since it is necessary, as the author says, to be able to go back in the opposite direction from the statement of the final article to the piece of forest to control the quality of the mediations thus created. Scientists spend their time carrying out these checks but « everything goes as if » Bruno Latour's narrative could not account for the rest of this work, while sparing himself the testing for his own work since it is a philosophical monograph and no digital traces collection could have been set up at this time.

*Ethnography* thus proceeds, by producing case studies, in cultural isolates whose dimensions can be reconstructed, and Latour'article fits entirely in this vein. On the other hand, the *anthropologist*'s task is to collect these field studies to make them comparable, as Mauss did, even though he had never set foot on "real" traditional social field. As the very form of the "essay on the gift" shows through the stacking of empirical references in footnotes that sometimes eat up all the text, Mauss proved to be an excellent comparatist. He did so not in the name of diversity that would have to be restored, which would add only noise and a claim to an unreachable "whole". His aim was to extract from all the observations a strong point, that of "hau" in this case. Mauss was therefore making radical reductions of all the ethnographic material available for argumentative reasons, from a particular point of view, which happened to be a standpoint on the agency of circulating entities. The strength of Bruno Latour's essay is of the same order in terms of the standpoint adopted. Farar from seeking to restore the "whole" of the field work of pedologists and botanists, he adopts a totally new and radically reducing view centred on the objects and the chain of information they circulate even though he would claim to repopulate the process of scientific reference, traditionally reduced to controversies about paradigms for instance in a kuhnina approachBut, despite the long description of the chain, it is still a reductionist view of the scientific work, it  si a ethnographic photograph of one specific team and not many, of a specific field work in the reserach program, of some specific tools and methods and not all of them. And there is no way escaping these constraints when doiing ethnography and sociography. However, the work fo the anthropologist of of the sociologist is to validate validate these statements, through careful comparison when the methods are qualitative and to calculate as much as possible this validation when indicators and data are set up. Unfortunately, after collecting this specific material from this specific field, Bruno Latour, unlike Mauss, could not rely on similar studies to make comparisons that would test the robustness of his hypothesis. Some readers might even consider this essay as an encouragement to the "thick" description, which must account for as many links as possible to gain credibility. B. Latour (1988) had already proposed such a horizon, when he wrote "Irréductions", which challenged the classical approaches in the social sciences.

## 2.3. Time for comparisons, reductions and computation has come for ANT and social sciences
ocol

However, Latour's own work could only be communicated at the price of constant reductions: in "laboratory life" (Latour and Woolgar, 1979), the photo of the laboratory roof does not lead him towards a thesis on laboratory architecture or fluid management principles, all important things in the maintenance of a "laboratory" entity, because he has other priorities This temptation of the "whole" which can be derivated from the expression "irreductions" is not supposed to be part of the ANT framework and was clearly challenged in our common paper on « the whole is smaller than its parts » (Latour,

Boullier and al. 2012), however not for the right reasons as it seems now. Because when ANT's method does not establish a protocol to make the comparison possible, it encourages the search for exhaustiveness or conversely the focusing on the "beautiful case", the beautiful « link » (the topofil for example) whereas another observer would have judged more « critical » the role of another device, like the pedocomparator, as we said before The philosophical ambition of Latour's paper enlargesthe constant gap between social science traditions and the work of Bruno Latour and of much of ANT. The absence of a testing protocol constitutes the greatest weakness of ANT among social sciences traditions, a weakness compensated by the richness and novelty of empirical observations, which are sufficient to change the reader's gaze (and to give him pleasure if the author is a good storyteller).

It is however possible to give a scientific status to this phase of exploration that constitutes this open observation, which aims to « repopulate » descriptions against the a priori reductionism of social sciences of the structure or of individual preferences.

It is therefore a *complementary phase* to this open exploration phase that should be put in place to validate the exploratory statements obtained by the ANT methods. Monitoring the actors, tracing the network of associations that are always unique by definition, cannot only depend on the implicit and indisputable choices of the observers, whose inspiring or sometimes theoretically correct approach is merely evaluated. In a second phase, it is possible to enter into a systematic comparison approach, provided one has a more assertive point of view, centred on what makes the specificity of ANT, namely the agency of the entities that circulate and weave this network that become the target of observations. But this requirement for comparability will necessarily end up reflecting on the methods of collecting information because monographs, despite their tendency to be exhaustive, may very well not capture some critical elements required for tracking

« replications » that become the central object of research. We need sociographic monographs indeed, provided that they can be reanalyzed by sociologists in order to compare them. However, the monographs and their materials themselves could be guided in a specific way depending on the standpoint that will be used later in the comparison. For instance, the transformation of a precise colour coding of soil samples (representational states) should be monitored and would take on more importance than the only account of the method to collect them (the device, the representational media).

For this monitoring, it is necessary to reach a degree of granularity that had hardly been possible until now, with qualitative and poorly equipped social science methods. The practice of science is changing thanks to digital infrastructures that fit into the entire reference chain. Hence, the observation of "science in the making" can also benefit from access to digital traces of a much finer grain. Showing the links and the role of each device remains essential but is still at a coarse grain if we really want to understand the translation process that allows information to emerge and be converted successively by the different links in the reference chain. For instance, the description of what the pedocomparator performorms is quite useful. However, the narrative does not provide the tables or codes that are generated by using this device, even though these representational states (tables, codes, figures and so on) will be necessary for the next step to design a diagram. The data that are vehicled require more detailed focus that the description of the representational media (the devices). And this attention on what circulates would be the only way to test whether these entities do have some agency, to the point where the devices and the human minds would be considered as the vehicles of the circulating representation. As Daniel Dennett said in a provocative way to demonstrate the radical shift of memetics **:** « A scholar is just a library's way of making another library. » (Dennett, 1990).These replications must be tracked down in order to account for the different chances of survival of these scientific data according to its properties (and according to the structure of the network or certain choices of propagators, if one adopts another standpoint). This evolutionary version of ANT, which would be coupled with Dennett's memetics, constitutes a research program that would make all these approaches more robust, thanks to digital traces A third generation of social sciences may find many insights in Dennett' works. He proposes to focus on « elements of culture » that live their lives through human minds. Their randopm variation are permanent, their combinations are an intrinsic part of human creativity but their selection can be accounted for and now digitally traces on specific platforms (social networks and all GAFAM), in scientific communities (as scientometrics demonstrates), in popular culture propagation (memes on Internet, Shifman, 2014) .

## 2.4. A tentative protocol for a replication-centered qualitative standpoint

If we sketch up the survey protocol that could now be put in place in the case of the Boa Vista topofil for comparative analysis, even calculable one if possible, here is the list of relevant tasks:

1/ Gather existing literature on the working methods of scientists (pedologists and botanists) as well as literature on their results.

2/ Build a comparative field with very similar cases on the domain itself (forest advancement).

3/ Carry out the field work for thick descriptions by restoring all the mediations as did B. Latour but in addition, collect the  digital traces that make it possible to trace the entities which circulate from one phase to the other allthrough the protocol. Make an inventory of the *representation states* that circulate (and not only of the devices, the *representational media*) that make the data emerge.

4/ Select certain data flows (here, « soil data » for example) and compare between the sites the methods of replication, i.e. both the forms of propagation but also the derivations, all the transformations that allow circulation. This is very similar to what Glaser and Strauss(1967) called « theoretical sampling », designed for generating concepts and theories.

 5/ Construct a new fieldwork which deals with other topics in pedology, then between different disciplines by focusing on replications of a certain type which allow comparisons (for example: the tranfer into diagrams, both from data collected directly but also from interviews with producers in order to understand their own work and their role as vehicles of these propagations, how they are affected by variations in formats, transmission rates or volumes, semiotic choices made, etc.). The shift is from topics (the issue actors are focusing on) to replications (the elements of the data that make the topic survive and the individual actors act). In doing so, we follow the tracks of Glaser and Strauss (1967) about « continuous comparison ». « While coding an incident for  a category, compare it with the previous incidents in the same and different groups coded in the same category ». They emphasized the need for reduction of terminology and consequent generalizing that are required to achieve a theory : « *parsimony* of variables and formulation and 2. *scope* in the applicability of the theory to a wide range of situations, while keeping a close correspondance of theory and data »  (p.111). The ANT tradition tends to favor reduction of terminology through the use of very large and versatile concepts (translation, attachments, assemblage) that allow extensive case-speciifc descriptions and rather loose comparisons, while we insist on the correspondance with data and the need for reduction of terminology in a more controllable way.

6/ If at least some of this data can be used for computer processing, collect also the datasets, including images (diagrams for instance) to launch a comparative processing on a larger scale, by further reducing the dimensions (example: some semiotic features of diagrams) without selecting them *a priori*.

In the end, do we gain anything from the previous statements already present in the original paper on « the topotfil of Boa Vista » ? still produce an accountof « science in the making », by reconstituting the mediations, but also by making them comparable, even calculable for a part of the properties which manufacture science. Individual researchers are no longer at the centre, nor are data collection devices, but above all everything that circulates between them. Of course, neither researchers nor devices nor situations are evacuated or ignored, but the focus on the data and signals that pass through them also makes it possible toexplore how data and messages i.e. replications are acting and making the next steps possible. « Science in the making » is also unfolding through the power of circulation of these statements, and small differences can now be traced, studied, even calculated to account for their specific agency.

## Conclusion

These kinds of protocols do not look as fancy and elegant as well written narratives. They require a long, collective and well equipped survey in order to make the comparison possible and the validation sound. In order to convince and to become more accountable, social sciences should move out of the monographic trend. Qualitative fieldwork needs to be more focused, at some moment in the process, around specific features, the ones that can be compared, which means reducing the focus and looking for thicker description limited to these features. However, the connection between qualitative and quantitative approach will be facilitated by these reductions since calculation means discretization, out of the analogous continuity, as we illsutrated in the case of a new protocol for the topofil of Boa Vista. This is where the digital capacities and especially the ones of the traces produced on the web and on social networks platforms make a decisive contribution. All the phenomena that can be traced at a very fine grained level can now be accounted for, paving the way for the new generation of social sciences, focused on propagation of elements of culture, on memes, on ideas, on signals and so on. It appears, as we tried to show in the case study of the topofil of Boa Vista, that this is especially adapted to the ANT principles where the agency is distributed to other entities than structures or individual preferences. However, this extension of ANT towards a more analytical stance, which is the condition for its connexion with the computable accounts, require to quit the irreductionist view of scientific activity that triggers some nostalgia of a « whole », through an extensive format of description of field studies. It does not mean that the requirements of thick description should be discarded but that it should be complemented by more comparable and formalized accounts and focused on the replications (the ones that circulate, propagate and make other actants act) that other social sciences used to underestimate and even to make disappear.

# References

Anderson, B. (1991). *Imagined Communities: Reflections on the Origin and Spread of Nationalism*.London and New York, Verso.

Barthes, R. (1957), *Mythologies*, Paris, Le Seuil.

Barthes, R. (1967), *Système de la mode*, Paris, Le Seuil.

Bessy C. et F . Chateauraynaud (1995). *Experts et faussaires. Une sociologie de la perception*, Paris, Métailié.

Blackmore, S. (1999), *The meme machine*, Oxford, Oxford University Press.

Blondiaux, L. (1998). *La fabrique de l'opinion. Une histoire sociale des sondages*. Paris: Le Seuil.

Boullier, D. & Lohard A. (2012). *Opinion mining et sentiment analysis. Méthodes et outils*. Paris: Open Editions Press.

Boullier, D. (2004). *La télévision telle qu'on la parle. Trois études ethnométhodologiques*. Paris: L'Harmattan.

Boullier, D. (2016). *Sociologie du numérique*. Paris: Armand Colin.

Boullier, D. (2010), *La ville-événement. Foules et publics urbains*, Paris, PUF.

Boullier, D. S. Chevrier et S. Juguet (2012), *Evénements et sécurité. Les professionnels des climats urbains*, Paris, Les Presses des Mines.

Boullier, D. (2017), « Big data challenges for social sciences: from society and opinion to replications », *ISA esymposium*, vol.7, issue n°2.

Boullier, D. (2018), Distribution du pouvoir d'agir des entités sociales dans les études informatiques sur Twitter, *Sociologie et Sociétés*, n°50 (in press).

Bourdieu P. (1984). Public Opinion Does Not Exist. In Mattelart A. and Siegelaub, S. (Eds.) *Communication and Class Struggle*. New York, International General.

Bourdieu P. (1979) *La distinction. Critique sociale du jugement*, Paris, Minuit.

Braudel F. (1958). Histoire et Sciences sociales: La longue durée. In: *Annales. Économies, Sociétés, Civilisations*. 13ᵉ année, N. 4, 1958. pp. 725-753.

Bromberger, C. (1995), *Le match de football. Ethnologie d'une passion partisane à Marseille, Naples et Turin*, Paris, Editions de la Maison des Sciences de l'Homme.

Callon M. (1979) «L'État face à l'innovation technique. Le cas du véhicule électrique» , *Revue Française de Science Politique*, XXIX, p.426-447

Cochoy, F. (1999). *Une histoire du marketing. Discipliner l'économie de marché*. Paris: La Découverte.

Dawkins, R. (1976), *The Selfish Gene*, Oxford, Oxford University Press.

Dennett, D. (2017), *From Bacteria to Bach and Back. The Evolution of Minds*, Penguin Books.

Dennett, D. ( 1990), Memes and the Exploitation of Imagination, *Journal of Aesthetics and Art Criticism*, 48, 127-35, Spring 1990.

Desrosières, A. (1998). *The Politics of Large Numbers: A History of Statistical Reasoning*. Cambridge Massachusetts: Harvard University Press.

Desrosières, A. (2014). *Prouver et gouverner: une analyse politique des statistiques publiques*. Paris: La Découverte.

Didier, E. (2002). "Sampling and Democracy: Representativeness in the First United States Surveys", *Science in Context*, 15 (3), 427-445.

Didier, E. (2009). *En quoi consiste l'Amérique? Les statistiques, Le New Deal et la démocratie*. Paris: La Découverte.

Douglas, J.D. (1967). *The Social Meanings of Suicide*. Princeton, NJ: Princeton University Press.

Durkheim, E. (1893). *The division of labour in society*, Paris, Alcan.

Durkheim, E. (1897). *Le suicide*. Paris, Alcan.

Durkheim, E. (1912). *Les formes élémentaires de la vie religieuse*. Paris, Alcan.

Eymard-Duvernay F., Favereau O., Orlean A., Salais R. & Thévenot L (2004). L'économie des conventions ou le temps de la réunification dans les sciences sociales. *Problèmes Économiques*, 2838.

Foucault, M. (1982). *The Archaeology of Knowledge and the Discourse on Language*. New York: Vintage Books.

Gallup, G. (1939). *Public Opinion in a Democracy*. Stafford: Herbert L. Baker Foundation Little Lectures.

Glaser, B. and A. Strauss (1967), *The Discovery of Grounded Theory*, Chicago, Aldine.

Hutchins E. (1995) *Cognition in the wild*, Cambridge, The MIT PressKatz, E. & Lazarsfeld, P. (1955). *Personal Influence: The Part Played by the People in the Flow of Mass Communication*. Glencoe: Free Press.

Latour, B (1987). *Science in Action, How to Follow Scientists and Engineers through Society.* Cambridge (Mass.): Harvard University Press.

Latour, B. (1995), « The 'Topofil' of Boa Vista-A Photo-Philosophical Montage », *Common Knowledge*, Vol. 4, n°1, pp 145-187.

Latour, B. (1996) « On interobjectivity », *Mind, Culture, and Activity: An International Journal.* Special symposium with discussion by Marc Berg Michael Lynch, Yrjo Engelström and a response by the author, Vol.3, n°4, pp.228-245 & 246-269.

Latour, B. (2005). *Reassembling the Social - An Introduction to Actor-Network-Theory*. Oxford: Oxford University Press.

Latour, B., Jensen B., Venturini T., Grauwin S. & Boullier D. (2012). The Whole Is Always Smaller Than Its Parts. A Digital Test Of Gabriel Tarde's Monads. *British Journal Of Sociology 63* (4), 590–615.

Latour B. (1996) 'On Interobjectivity », *Mind, Culture, and Activity: An International Journal* Special symposium with discussion by Marc Berg Michael Lynch, Yrjo Engelström and a response by the author, Vol.3, n°4, pp.228-245 & 246-269.

Latour B. (1988) *The pasteurization of France*, Harvard University Press, Cambridge Mass

Mauss, M. (1950). *Essai sur le don. Sociologie et Anthropologie*, Paris, Puf.

Mc Luhan, M. (1964). *Understanding Media: The Extension of Man*, London, Routledge.

Osborne, T. & Rose, N. (1999). Do the social sciences create phenomena? The example of public opinion research. *British Journal of Sociology 50*(3), 367-396.

Pinch, T. J. and W. E. Bijker (1984), "The Social Construction of Facts and Artefacts : or How the Sociology of Science and the Sociology of Technology might Benefit Each Other", *Social Studies of Science*, Vol. 14.

Preda, A. (2006), « Socio-Technical Agency in Financial Markets. The Case of the Stock Ticker », *Social Studies of Science*, Volume: 36 issue: 5, page(s): 753-782.

Tarde, G. (1890). *Les lois de l'imitation*. Paris, Alcan.

Tarde, G. (1901). *L'opinion et la foule*. Paris, Alcan.

Tarde, G. (1897), *Contre Durkheim. A propos de son « suicide »,* édité par P. Besnard et M. Borlandi, Université du Québec à Chicoutimi.

Tönnies, F. (1887). Gemeinschaft und Gesellschaft, Leipzig, Fues's Verlag.

Watts, D.J.& P.S. Dodds (2007). Influentials, networks, and public opinion formation. *Journal of Consumer Research*, vol. 34, no. 4, 441–458.